\documentclass[aps,pre,showpacs,twocolumn,superscriptaddress]{revtex4-1}
\usepackage{graphicx}
\usepackage{txfonts}
\usepackage{amssymb}
\usepackage{makeidx}
\usepackage{bbold}

\def\dd{\mathrm{d}}

\def\simge{\mathrel{%
    \rlap{\raise 0.511ex \hbox{$>$}}{\lower 0.511ex \hbox{$\sim$}}}}
\def\simle{\mathrel{
    \rlap{\raise 0.511ex \hbox{$<$}}{\lower 0.511ex \hbox{$\sim$}}}}

\newcommand \be{\begin{eqnarray}}
\newcommand \ee{\end{eqnarray}}
\newcommand{\del}{\partial}

\def\XXint#1#2#3{{\setbox0=\hbox{$#1{#2#3}{\int}$}
\vcenter{\hbox{$#2#3$}}\kern-.5\wd0}}

\begin{document}

\title{Universal shocks in the Wishart random-matrix ensemble - a sequel}

\author{Jean-Paul Blaizot}
\email{Jean-Paul.Blaizot@cea.fr} \affiliation{IPTh,  CNRS/URA 2306, CEA-Saclay,
91191 Gif-sur Yvette, France}

\author{Maciej A. Nowak}
\email{nowak@th.if.uj.edu.pl} \affiliation{M. Smoluchowski Institute
of Physics and Mark Kac Center for Complex Systems Research,
Jagiellonian University, PL--30--059 Cracow, Poland}

\author{Piotr Warcho\l{}} \email{piotr.warchol@uj.edu.pl} \affiliation{M.
Smoluchowski Institute of Physics,  Jagiellonian University,
PL--30--059 Cracow, Poland}
\date{\today}

\begin{abstract}
We study the diffusion of complex Wishart matrices and derive a partial differential equation governing 
 the behavior of the associated averaged characteristic  polynomial. In the limit of large size matrices, the inverse Cole-Hopf transform of this polynomial  obeys a nonlinear 
partial differential equation whose solutions exhibit shocks at the evolving edges of the eigenvalue spectrum. In a particular 
scenario one of those shocks hits the origin that plays the role of an impassable wall. To investigate the universal behavior in the vicinity of  this wall, a critical point, we derive an integral representation for the 
averaged characteristic polynomial and study its asymptotic behavior. The result is a Bessoid function.

\end{abstract}

\maketitle

\section{Introduction}

The Wishart random matrix ensemble \cite{WISHART}, a multidimensional generalization of the $\chi$-squared distribution, has
proven, over the many
years since its
invention, to be one of the most prominent examples of the vast applicability of random matrix theory. It has become an important
tool in multivariate
statistics
\cite{SILVERSTEIN}, helping to understand a broad range of phenomena occurring in such fields as population structure study
\cite{GENETICS}, financial
data analysis
\cite{BOUCHPOTT} or image processing \cite{IMAGE}. When it was realized
that it can describe the information capacity of a multiple input multiple output system \cite{FOSCHINI,DEBBAH,TELATAR},
the otherwise called Laguerre ensemble, has changed the face of multichannel information theory. Moreover,
being closely related to so called
chiral random matrices, the Wishart matrix shares, in a narrow, universal window in the vicinity of the zero eigenvalue, spectral
properties with the
Dirac operator in
Euclidean Quantum Chromodynamics, thus portraying the spontaneous breakdown of chiral symmetry through the famous Banks-Casher
formula
\cite{VERBAARSCHOTREVIEW}. Finally,
matrices from the Laguerre ensemble appear in quantum information theory \cite{KAROL}, the research of conducting mesoscopic systems
\cite{CARLO} or chaotic scattering in cavities \cite{CHAOS}.

The study of static properties of random matrices proves to be highly rewarding. 
Yet, as realized already by Dyson \cite{DYSON},
introducing some additional dynamics can be equally if not more fruitful. In the case of the Wishart ensemble, an evolving matrix
was first defined
through a Brownian
motion of real and complex matrix entries in \cite{BRU, BRU2} and \cite{AKUWAD, KONIG}, respectively. More recently \cite{VIVO},
such a stochastic
process was generalized
to arbitrary values of the Dyson parameter $\beta$, in particular for $\beta\in (0,2]$. In the mean time, the theory of
non-intersecting Brownian
motions or the so called
vicious walkers was developed. The subject which originated from the works of de Gennes on fibrous structures \cite{DEGENNES},
and Fisher on wetting
and melting
\cite{FISHER}, was linked to random matrix theory \cite{RMTVW, JOH, KATORI, GREGORYSATYA}, and led to many developments including
a physical
realization of the
statistical properties of Wishart matrices through fluctuations of non-intersecting interfaces in thermal equilibrium
\cite{CELINESATYA}.

Both in random matrix and vicious walker theories, a central role is played by (multi-)orthogonal polynomials, their Cauchy
transforms and the related,
characteristic and
inverse characteristic polynomials. This is because these polynomials are the building blocks of correlation functions, and they govern the universal
asymptotic behavior
of probability
distributions \cite{FYOSTRA1, FYOSTRA2, KUIJLAARS1, DELVAUX}. It is an ongoing challenge to uncover the properties of these
objects, in particular,
those related to the
Laguerre ensemble.

In a previous work \cite{BNW1}, we have studied the stochastic evolution of a Wishart matrix for trivial initial conditions corresponding to vanishing eigenvalues. In that
setting, the
associated characteristic
polynomial coincides with a time dependent, monic, orthogonal Laguerre polynomial which we have shown to satisfy a certain, exact
(i.e., it is valid for any 
matrix size $N$),
complex, partial differential equation. This, in turn, allowed us to recover the universal Airy and Bessel asymptotic behaviors of
the characteristic
polynomial at the
edges of the spectrum as associated with hydrodynamical like shocks arising from the solution of a related nonlinear partial
differential equation
governing the evolution
of the resolvent in the large $N$ limit.

Here, we show that the same stochastic process, but with non trivial initial conditions, i.e. initialized with a Wishart matrix possessing a single $N$-degenerate
eigenvalue $a^2\ne 0$, allows us 
to identify  a
microscopic eigenvalue scaling associated with a novel asymptotic behavior of the characteristic polynomial. The phenomenon occurs
at the origin,
precisely when it is hit
by the diffusing spectrum or, in the hydrodynamic language of \cite{BNW1}, when the shock wave reaches the origin that plays the role of an impassible wall. To achieve
this, we prove that
the characteristic
polynomial satisfies the above mentioned partial differential equation for any initial condition. In this scenario, the model can be viewed as a Wishart matrix perturbed by a source and there are no polynomials, orthogonal in the classical sense, associated with this setting.
Note that this is the reason why the derivation requires the use of more sophisticated methods than
 those employed in \cite{BNW1}. 
 Moreover, it was through the studies of the Gaussian unitary random matrix ensemble with an 
external source \cite{PZJ1, PZJ2} that the asymptotic Pearcey behavior at the critical point was
 discovered \cite{BREZINHIKAMIGAP, TRACYWIDOMP, BLEHERKUIJLAARS}. In our setting, it would arise through a diffusion of a Hermitian matrix initiated with at least two distinct eigenvalues, at the point of merging of the spectra.  
In the case of the Laguerre ensemble, the additional symmetry imposes a different functional form, associated with modified Bessel functions of the first kind.

The multiple orthogonal polynomials associated with the modified Bessel functions of the first kind were first studied in \cite{COAS1} and \cite{COAS2}.
They were used to build a kernel for a chiral Gaussian unitary ensemble perturbed by a source in \cite{FORRESTER}. The critical behavior studied here was however
not identified. This was done in the context of non-intersecting squared Bessel paths in \cite{KMFW} where the integral representation
 of the limiting kernel was derived with Riemann-Hilbert techniques. Finally, this kernel reduces to the so-called symmetric Pearcey kernel
 identified through the studies of random growth with a wall  \cite{BOKU, KUAN}. Our work differs from those above by the use of completely different methods. We follow strictly the diffusing Wishart matrix and focus on the averaged
 characteristic polynomial rather than on the kernel itself. This allows us to obtain a unified picture  of the behavior near the critical point. 

This paper has the following structure. We start by defining the Brownian walk of the elements of a Wishart matrix and state the
partial differential
equations fulfilled
by the associated characteristic polynomial and its logarithmic derivative - an inverse Cole-Hopf transform (the
proof is left for the
appendix). The
latter coincides with the resolvent (or Green's function) in the large $N$ limit. By  exploiting the method of complex characteristic, we subsequently
determine the resolvent
for our new set of
initial conditions. This allows us to formally identify the new critical point and the large $N$ scaling of the eigenvalue density
in its vicinity. We
then recover 
the explicit, arbitrary $N$ solution of the partial differential equation for the characteristic polynomial, expand it in the
vicinity of the origin
at the time of the
collision and show that it is asymptotically a version of the so-called Bessoid function. We end the paper with conclusions.

\section{Formal setting}
We consider a $N\times N$ random matrix of the following form:
\be
L(\tau)=K^{\dagger}(\tau)K(\tau).\label{chiral1}
\ee
where the entries of $K$, an $M\times N$ ($M>N$) matrix,  evolve in time $\tau$ 
according to
\be
{\rm d}K_{ij}(\tau)=\dd x_{ij}+i\dd y_{ij}=b^{(1)}_{ij}(\tau)+ib^{(2)}_{ij}(\tau),\label{diff0}
\ee
where $b^{(1)}_{ij}(\tau),b^{(2)}_{ij}(\tau)$ are two independent sets of free Brownian walks: 
\be 
b^{(c)}_{ij}(\tau)=\zeta^{(c)}_{ij}(\tau){\rm d}\tau,\ee \be\left\langle \zeta^{(c)}_{ij}(\tau)\right\rangle = 0
\ee 
and 
\be
\left\langle \zeta^{(c)}_{ij}(\tau)\zeta^{(c')}_{kl}(\tau')\right\rangle
=\frac{1}{2}\delta^{cc'}\delta^{ik}\delta^{jl}\delta(\tau-\tau').
\ee 
We define  $\nu\equiv M-N$ and the rectangularity as $r\equiv N/M$. 

To the free Brownian motions is associated a (Gaussian) probability which allows us to define the averaged characteristic polynomial associated with the matrix $L$: 
\be\label{Mdef}
Q_{N}^{\nu}(z,\tau)\equiv\left\langle\det\left[z- L\right]\right\rangle.
\ee
It is shown in 
appendix A that 
$Q_{N}^{\nu}(z,\tau)$ satisfies the following partial differential equation
\be
\del_{\tau}Q_{N}^{\nu}(z,\tau)=-z\del_{zz}Q_{N}^{\nu}(z,\tau)-(\nu+1)\del_{z}Q_{N}^{\nu}(z,\tau)\label{pde1}
\ee
for any initial condition. The same equation was obtained in \cite{BNW1} for the particular initial condition  $L(\tau=0)=0$. It was shown there, that its solutions are in this case the time dependent, monic, Laguerre polynomials.

We proceed by performing the inverse Cole-Hopf transform on $Q^{\nu}_{N}(z,\tau)$. Namely, we define
$f_{N}=\frac{1}{N}\del_{z}{\rm
ln}(Q_{N}(z,\tau))$. Eq.~(\ref{pde1}) then yields the following equation for $f_{N}$:
\be 
\del_{\tau}f_{N}+2Nzf_{N}\del_{z}f_{N}+N f_{N}^{2}=-(2+\nu)\del_{z}f_{n}-z\del_{zz}f_{N}.\label{ }
\ee
After rescaling the time according to $\tau\to\frac{r\tau}{N}$ \cite{BN2,BNW1}, this equation becomes 
\be \nonumber
\del_{\tau}f_{N}+r\left(2zf_{N}\del_{z}f_{N}+f_{N}^{2}\right)+(1-r)\del_{z}f_{N}\\=-\frac{r}{N}\left(2\del_{z}f_{N}+z\del_{zz}f_{N}\right).
\label{eqf2}\ee
In the large $N$ limit, $f_{N}(z,\tau)=G(z,\tau)\equiv\frac{1}{N}\left<\rm Tr\frac{1}{z-L(\tau)}\right>$, we recover:
\be \nonumber
\partial_{\tau}G(z,\tau)=(r-1)\partial_z G(z,\tau)+\\-2rzG(z,\tau)\partial_z G(z,\tau)-rG^2(z,\tau),
\label{wishartburgers2}\ee
in agreement with ~\cite{DUGU}. The  partial differential equations (\ref{pde1}) and (\ref{wishartburgers2}) form the backbone of
this paper. Solving
the later, in the
following section, will allow us to recover the large $N$ limit spectrum of eigenvalues and identify  the scaling of
the level  density in the
vicinity of the
edges, in particular that near the origin. The former, on the other hand, as shown in the fourth section, admits an asymptotic solution that describes the universal behavior near the critical point. This new solution is the main result of this paper.

\section{The bulk of the spectrum at large $N$}
As announced above, we choose  for the initial condition $L(\tau=0)=\mathbb{1}^{N\times N}a^2$. This, according to Eq.~(\ref{Mdef}), translates into
$Q_{N}^{\nu}(z,\tau=0)=(z-a^2)^{N}$
and, in the large $N$ limit, 
$G(z,\tau=0)=\frac{1}{z-a^2}$. We focus in this section on $G(z,\tau)$, and solve Eq.~(\ref{wishartburgers2}) using the  method of complex characteristics. This method transforms
Eq.~(\ref{wishartburgers2}) into
the following three
ordinary differential equations \cite{BNW1}:
\be\frac {{\rm d}z}{{\rm d}s}=1-r+2rzG,\label{ode3}\ee
\be\frac {{\rm d}\tau}{{\rm d}s}=1\label{ode1},\ee
\be\frac {{\rm d}G}{{\rm d}s}=-r{G}^{{2}}\label{ode2},\ee
such that $z(s=0)=z_0+a^{2}$, $\tau(s=0)=0$, and $G(s=0)=\frac{1}{z_{0}}$. Solving the last two equations yields $s=\tau$ and
\be\label{Grtau} G=\frac{1}{r\tau+z_{0}}.\label{G1}\ee 
We are therefore left with:
\be
\frac{{\rm d}z}{{\rm d}\tau}=1-r+\frac{2rz }{r\tau+z_{{0}}},
\ee
which is solved by:
	\be z=\left({{z}_{{0}}+r\tau}\right)\left(1+\frac {\tau}{z_{0}}+a^{2}\frac{\tau r+z_0}{z_{0}^{2}}\right)\label{char1}.\ee
The characteristic curves are parameterized by $z_0$. By eliminating $z_0$ in Eq.~(\ref{G1}) one gets the following, implicit, cubic equation for $G(z,\tau)$:	
\be \label{Gimplicit} z=\frac{1}{G(z,\tau)}+\frac{\tau}{1-r\tau G(z,\tau)}+a^{2}\frac{1}{\left(1-r\tau
G(z,\tau)\right)^{2}}\label{green1}.\ee
The proper  solution of this equation yields the eigenvalue density via the usual Sochocki$-$Plemelj formula. An illustration of this density, and its time-dependence, is given by Fig.~\ref{rho}. One
can also reconstruct
the spectrum directly from the characteristic curves, as shown in appendix B.
	\begin{figure}[htbp]
		\centering
		\includegraphics[width=0.44\textwidth]{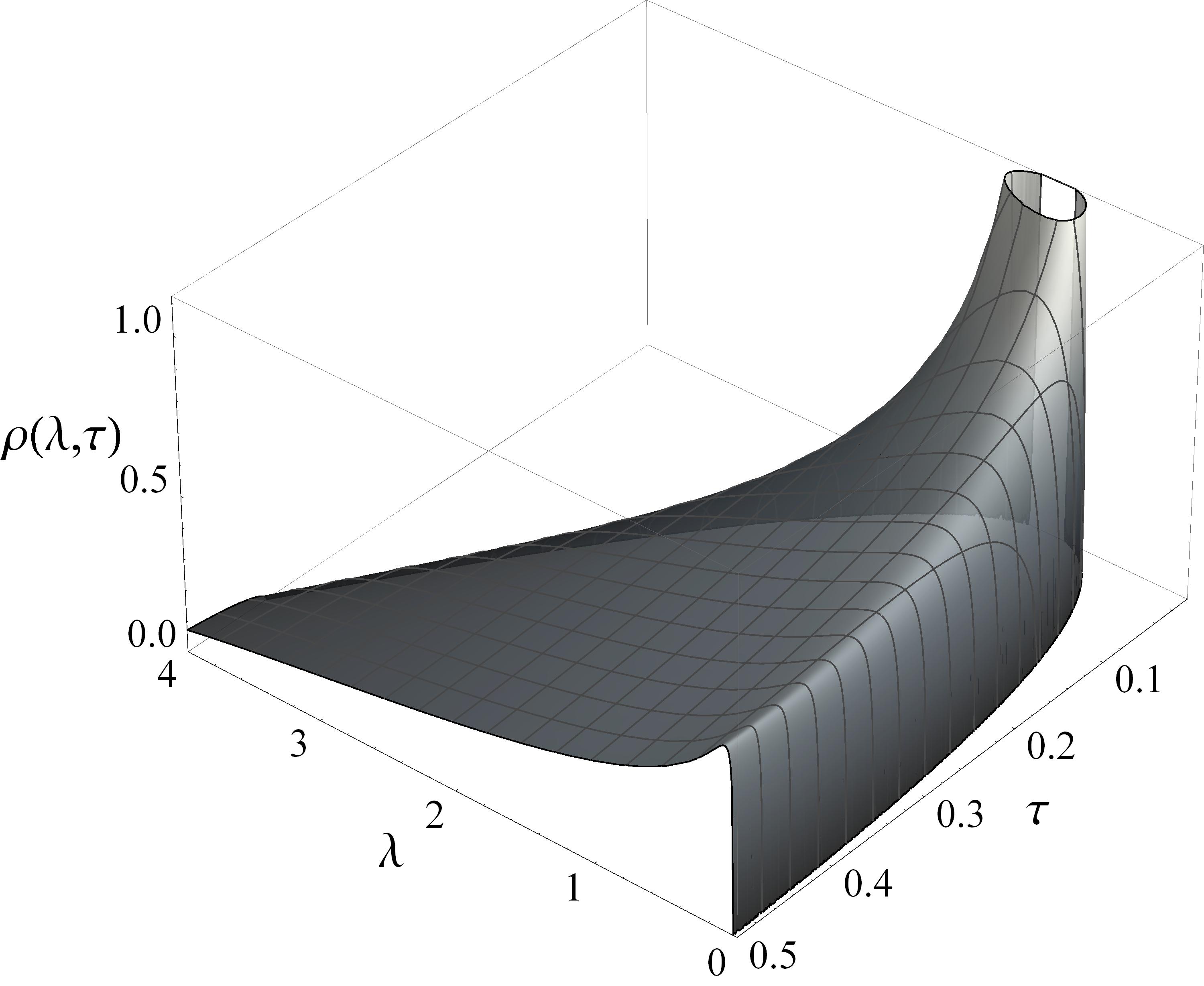}
		\caption{The computed eigenvalue probability density function for $a=1$ and $r=1$. }
		\label{rho}
	\end{figure}

From now on we will work in the $r=1$ limit, since only then the eigenvalues can reach the origin. This is realized when we let $N$ and $M$
go to infinity keeping $\nu$
 constant and finite.

In the above derivation we assumed that the mapping between $z$ and $z_0$ is one-to-one, that is, it can be inverted. This is the
case except at points
$z_{0_{c}} (\tau)$,
such that ${\rm d}z/{\rm d}z_{0}=0$, where a singularity occurs. We obtain the following equation for $z_{0_{c}}$:
\be
z_{0c}^{3}-z_{0c}\tau(2a^2 +\tau)-2a^ 2\tau^2 =0.\label{shock1} 
\ee
The equation  defines  the location of the shock waves, which coincide with the edges of the
spectrum. From this equation, we
deduce that the shock
wave reaches $0$ at $\tau_{c}=a^2$.

In the vicinity of the critical point $(z=0, \tau=a^2)$,  we have, in the leading order ($s$ small and $\tau$ near $a^2$):
\be
G(s, \tau)-\frac{2}{3\tau}\propto s^{-\frac{1}{3}}.  
\ee
This yields the average eigenvalue spacing scaling as $N^{-3/2}$.
If we further allow
ourselves to move
around the critical point within the time domain, a careful expansion of $G(s, a^2 +t)$ will show that $t$ has to be of the order
of $N^{-1/2}$.
In the beginning of the paper we have defined the evolution of the matrix elements $K_{ij}$  so that $\left\langle |K_{ij}|^2\right\rangle\sim\tau$. 
The time is rescaled overall by $N^{-3/2}$ and therefore this diffusive character of the dynamics is preserved in the microscopic regime defined in the vicinity of the critical point.   

We are now equipped with enough information to study, in the following section, the large $N$, $M$ asymptotics of the
characteristic polynomial.


\section{The characteristic polynomial at the critical point}

Recall the partial differential equation for the averaged characteristic polynomial (\ref{pde1}): 
\be
\del_{\tau}Q_{N}^{\nu}(z,\tau)=-\frac{1}{M}z\del_{zz}Q_{N}^{\nu}(z,\tau)-\frac{(\nu+1)}{M}\del_{z}Q_{N}^{\nu}(z,\tau)\label{pde2}.\ee
We found a solution for any $N$ and any $\nu$ of this equation. It reads:
\be\nonumber
Q_{N}^{\nu}(z,\tau)= \mathcal{C} \,\tau^{-1} z^{-\frac{\nu}{2}} \int^{\infty}_{0} y^{\nu +1}\\
\exp{\left(M\frac{z-y^{2}}
{\tau}\right)}I_{\nu} \left(\frac{2iM y}{\tau}\sqrt{z}\right)Q^{\nu}_{N}(-y^2,0)\dd y,
\ee
where $I_{\nu}$ is the modified Bessel function. The constant $\mathcal{C}$ is found by matching the solution with the initial condition $Q^{\nu}_{N}(z,0)$. 
Note that \cite{AS} $\lim_{|x|\to\infty}I_{\nu}(x)\simeq\frac{1}{\sqrt{2\pi
x}}e^{x}$, valid for $|\arg(x)|<\frac{\pi}{2}$ and here $x=\frac{2iMy\sqrt{z}}{\tau}$ so that $\arg{(z)}\ne 0$. 
In the limit of $\tau\to 0$, the saddle point approximation method enables us to deduce that $\mathcal{C}=i^{-\nu}2 M $.  
We therefore obtained an integral representation for the averaged characteristic polynomial associated with a freely diffusing Wishart type matrix of arbitrary size 
and for arbitrary initial conditions consistent with the symmetry of the ensemble.  Let us mention, additionally, that it has been recently derived in \cite{PFOR}, with combinatorial methods, for a static Wishart matrix perturbed by a source.

We now turn to the specific case of $Q^{\nu}_{N}(z,0)=\left(z-a^{2}\right)^{N}$. In the limit of $M$ and $N$ going to infinity, with $\nu$ constant, the exponent, 
arising in the integral from the expansion of the modified Bessel function and the exponentiation of the initial condition,
 is dominated by values of $y$ in the vicinity of the saddle points given by the solutions of the equation:
\be
y-i\sqrt{z}-\frac{\tau y}{a^2+y^2}=0.
\ee
The three saddle points merge at $y=0$ for $z=0$ and $\tau=a^2$. Moreover, as predicted in the previous section, the critical behavior 
occurs when $|z|\sim N^{-3/2}$ and $|\tau-a^2|\sim N^{-1/2}$. One can therefore expand the natural logarithm $\ln(a^{2}+y^{2})\approx
\ln(a^{2})+\frac{y^{2}}{a^{2}}-\frac{y^{4}}{2a^{4}}$. Furthermore we set $\tau=a^{2}+N^{-1/2}a^2 t$ and $z=N^{-3/2}a^2 s$, with $\arg{(s)}\ne 0$. 
To recover the proper asymptotics we rescale the integration variable by defining $y=N^{-1/4}a u$. The limiting behavior becomes 
\be\nonumber
Q_{N}^{\nu}\left(N^{-\frac{3}{2}} a^2 s,a^{2}+N^{-\frac{1}{2}}a^{2}t\right)\approx
(-a^2)^N N^{\frac{\nu+1}{2}}s^{-\frac{\nu}{2}}\\
\int_{0}^{\infty}u^{\nu +1}\exp{\left(-\frac{1}{2}u^4+u^{2}t\right)}I_{\nu}\left(2iu\sqrt{s}\right)\dd u, \label{bessoid1}
\ee
the announced result. 

Let us mention that for $\nu=-\frac{1}{2}$, (\ref{bessoid1}) takes the form of:
\be
(i\pi)^{-\frac{1}{2}}s^{-\frac{1}{4}}\int_{0}^{\infty}\exp{\left(-\frac{1}{2}u^4+u^{2}t\right)}\cos\left(2u\sqrt{s}\right)\dd u
\ee
and is called the symmetric Pearcey integral through its connection with the symmetric Pearcey kernel arising for phenomena of random surface growth with a wall \cite{BOKU}.

Moreover, for positive integer $\nu$, as the Wishart ensemble is connected to Chiral random matrices, it has an analog in the 
integral describing the statistical properties of the Dirac operator around its zero eigenvalue, at the moment of chiral symmetry breaking in 
Euclidean Quantum Chromodynamics \cite{CHIRBURG}. The averaged characteristic polynomial of a diffusing complex chiral matrix is namely defined by
\be
\tilde Q^{\nu}_{M+N}(w,\tau)\equiv\left\langle{\rm det}\left(\begin{array}{cc} w & -K^{\dagger} \\ -K & w \end{array}\right)\right\rangle
\ee
and related to its Wishart counterpart through $\tilde Q^{\nu}_{M+N}(w,\tau)=w^\nu\,Q_{N}^{\nu}(z=w^2,\tau)$. Its critical point analysis is analogical.

Finally, (\ref{bessoid1}) was known earlier in optics. In particular
\be
B(x,y)\equiv \int_{0}^{\infty}u\exp{\left(iu^4+iu^{2}y\right)}I_{0}\left(i u x\right)\dd u,
\ee
is recognized as the Bessoid canonical function of order zero and appears in the description of the rotationally symmetric cusp (cuspoid) diffraction catastrophe~\cite{NYE,KA,BJ}. Note that the 
behavior of the two differ as the $\sqrt{s}$ is complex while $x$ is real and because the exponent in the latter has a complex phase.
We are however inspired by the analogy and call (\ref{bessoid1}) simply, the Bessoid function.

%
\section{Conclusions}
In this paper we have continued our study of matrices belonging to the Wishart ensemble and performing a white noise driven Brownian walk. 
Our derivation of the partial differential equation fulfilled by the associated averaged characteristic polynomial allows to inspect this process 
for arbitrary initial conditions and size of the matrix $N$. Here, we differ from the prequel of this paper, where the method used permitted only a study of 
a trivial initial condition, for which the average 
characteristic polynomial coincides with a Laguerre polynomial.
 
The inverse Cole-Hopf transform of the characteristic polynomial obeys a nonlinear partial
differential equation. In the large matrix size limit its solutions contain shocks which are positioned at the moving edges of the spectrum.
For a matrix diffusion initiated form a non-zero $N$ degenerate eigenvalue, when the shock reaches the origin, a distinct, universal behavior
 of the eigenvalues occurs in a window shrinking like $N^{-3/2}$ while $N$ grows to infinity. This phenomena is encapsulated by the asymptotics 
of the averaged characteristic polynomial. We have derived its integral representation and studied it in the vicinity of the critical point.
The resulting limiting behavior is described by an integral that we call Bessoid function.

\section*{Acknowledgments}
PW would like to thank the organizers of the 2012 Les Houches school on random matrices and integrable systems, where part of this
work has been done. PW is supported by the International PhD Projects Program
of the Foundation for Polish Science within
the European Regional Development Fund of the European
Union, agreement no. MPD/2009/6. MAN is  supported in part by the Grant DEC-2011/02/A/ST1/00119 of the National Center of Science.

\appendix
\section{Derivation of the PDE}
Here we derive the partial differential equation governing the evolution of the averaged
 characteristic polynomial associated with a diffusing Wishart matrix (\ref{pde1}). 
The real and imaginary parts of each of the elements of the matrix $K$ evolve according to the same diffusion equation:
\be\nonumber
\frac{\dd}{\dd \tau}P^{(1)}_{ji}=\frac{1}{4}\frac{\dd^{2}}{\dd x_{ji}^{2}}P^{(1)}_{ji}\\
\frac{\dd}{\dd \tau}P^{(2)}_{ji}=\frac{1}{4}\frac{\dd^{2}}{\dd y_{ji}^{2}}P^{(2)}_{ji}.\ee
The initial conditions are arbitrary. If however, we intend to stay in the realm of Wishart type random matrices, they cannot violate the
symmetry of the ensemble.
Since the elements evolve independently, the joint probability density is $P\left(x,y,\tau\right)=\prod_{j,i,c}P_{ji}^{(c)}$ and it obeys the equation
\be
\del_{\tau}P\left(x,y,\tau\right)=\frac{1}{4}\sum_{j,\,i}(\del_{x_{ji}x_{ji}}+\del_{y_{ji}y_{ji}})P\left(x,y,\tau\right).\label{diff}
\ee

Let $\eta$ represent a column of complex Grassman variables $\eta_i$ where $i\in\{1,2,...,N\}$.  
The averaged characteristic polynomial, associated with (\ref{chiral1}), can be
expressed in terms of the following integral :
\begin{widetext}
\be Q_N^\nu (z,\tau)=\int {\rm D}(\eta, \bar{\eta}, x, y)\; \exp{\left[\eta^\dagger \left(z-K^{\dagger}K\right)\eta \right]}P\left(x,y,\tau\right),
\ee
where the integration measure is ${\rm D}(\eta,\bar{\eta}, x, y)\equiv\prod_{i,j,k} {\rm d}{\eta}_{k}{\rm d}\bar{\eta}_{i}{\rm d}x_{ji}{\rm d}{y}_{ji}$.
This form allows us to proceed to the main part of the proof.
Acting with the time derivative on $Q_{N}^{\nu}(z,\tau)$ and exploiting (\ref{diff}), yield
\be
\del_{\tau}Q_{N}^{\nu}(z,\tau)=\frac{1}{4}\int{\rm D}(\eta,\bar{\eta}, x, y)\; \exp{\left[\eta^\dagger \left(z-K^{\dagger}K\right)\eta \right]}
\sum_{j,\, i}(\del_{x_{ji}x_{ji}}+\del_{y_{ji}y_{ji}})P\left(x,y,\tau\right).
	\ee	
At this point, we integrate by parts and proceed with the differentiation with respect to $x_{ji}$ and $y_{ji}$. After a brute force calculation one obtains
\be
\del_{\tau}Q_{N}^{\nu}(z,\tau)=-\int {\rm D}(\eta,\bar{\eta}, x, y)\;\eta^{\dagger}\eta
\left( M+\eta^\dagger K^{\dagger}K\eta\right)\exp{\left[\eta^\dagger \left(z-K^{\dagger}K\right)\eta \right]}
P\left(x,y,\tau\right) .
\ee
The first term in the sum can be represented as a differentiation with respect to $z$, the second one as a differentiation over Grassmann variables:
\be
\del_{\tau}Q_{N}^{\nu}(z,\tau)=-M\del_{z}Q_{N}^{\nu}(z,\tau)
+\int{\rm D}(\eta,\bar{\eta}, x, y)\; \eta^{\dagger}\eta \exp{\left(\eta^\dagger\eta z\right)} \sum_{i}\overline{\eta}_i \del_{\overline{\eta}_i }
\exp{\left(-\eta^\dagger K^{\dagger}K\eta \right)}
P\left(x,y,\tau\right) .
\ee
Again, we integrate by parts, this time in the Grassmann variables, and perform the differentiation. The result is

\be
\del_{\tau}Q_{N}^{\nu}(z,\tau)=-z\del_{zz}Q_{N}^{\nu}(z,\tau)+(N-M-1)\del_{z}Q_{N}^{\nu}(z,\tau),
\ee
which concludes the proof.
\end{widetext}

\section{Analysis of the characteristics}
Here,  we show how the large $N$ properties of the eigenvalue spectrum are encoded in the characteristics.
The complex characteristic curves are defined in the $(z,\tau)$ hyperplane by Eq.~(\ref{char1}), namely:
\be z=\left({{z}_{{0}}+r\tau}\right)\left(1+\frac {\tau}{z_{0}}+a^{2}\frac{\tau r+z_0}{z_{0}^{2}}\right)\label{char2}.\ee
These are labeled by the values of the complex variable $z_0$. They are not straight lines as in the case of the usual Burgers equation.
Let us define
$z=\lambda+i\eta$ and
$z_0=x+iy$. Notice that for $\tau=0$, $z=z_0$, so that if a characteristic starts from a purely real point $z_0$, then $z$  remains real at all times. For simplicity we set $a=1$ and, as we are interested in the  scenario where the spectrum hits the origin, $r=1$. By taking the real and the imaginary parts of Eq.~(\ref{char2}) one gets
\be
\lambda =\frac{2 \tau^2 x^2}{\left(x^2+y^2\right)^2}+\frac{\tau (\tau (x-1)+2 x)}{x^2+y^2}+2\tau+x+1\label{lambda},
\ee and
\be
\eta =y \left(1
-\frac{2 \tau^2 x}{\left(x^2+y^2\right)^2}-\frac{(\tau+2)\tau}{x^2+y^2}\right)\equiv y Y(x,y).\label{eta}
\ee 
We also have  (cf. Eq.~(\ref{Grtau}))
\be
G=\frac{1}{\tau+z_0}=\frac{\tau+x-iy}{(\tau+x)^2 +y^2}.
\ee
Moreover, the spectral density is given by $\rho(\lambda, \tau)=-\frac{1}{\pi}\Im G|_{\eta=0^{+}}$, and therefore 
\be
\rho(\lambda, \tau)=\left. \frac{y}{\pi\left[(\tau+x)^2 +y^2\right]}\right|_{\eta=0^+},
\ee
where $\lambda$ appears on the right hand side through (\ref{lambda}) and (\ref{eta}).
The limit $\eta=0^{+}$ can be accessed in four ways, by $y\to 0^{+}$ with $Y(x,y)>0$, $y\to 0^{-}$ with $Y(x,y)<0$
and through $Y(x,y)\to 0^{+}$ with $y>0$ or $Y(x,y)\to 0^{-}$ with $y<0$.
The first two give zero spectral density in the area of the characteristics curves defined by
\be
\lambda=\left(x+\tau\right)\left(1+\frac{\tau}{x}+\frac{\tau +x}{x^{2}}\right).
\ee
These are the curves which remain on the plane of real $z$ throughout the evolution.

The last two conditions, together with Eq.~(\ref{lambda}) define the characteristic curves which cross the $\eta=0$ plane at a
specific time $\tau$,
and reconstruct the
nonzero part of the spectral density. 

\begin{figure}[htbp]
		\centering
		\includegraphics[width=0.39\textwidth]{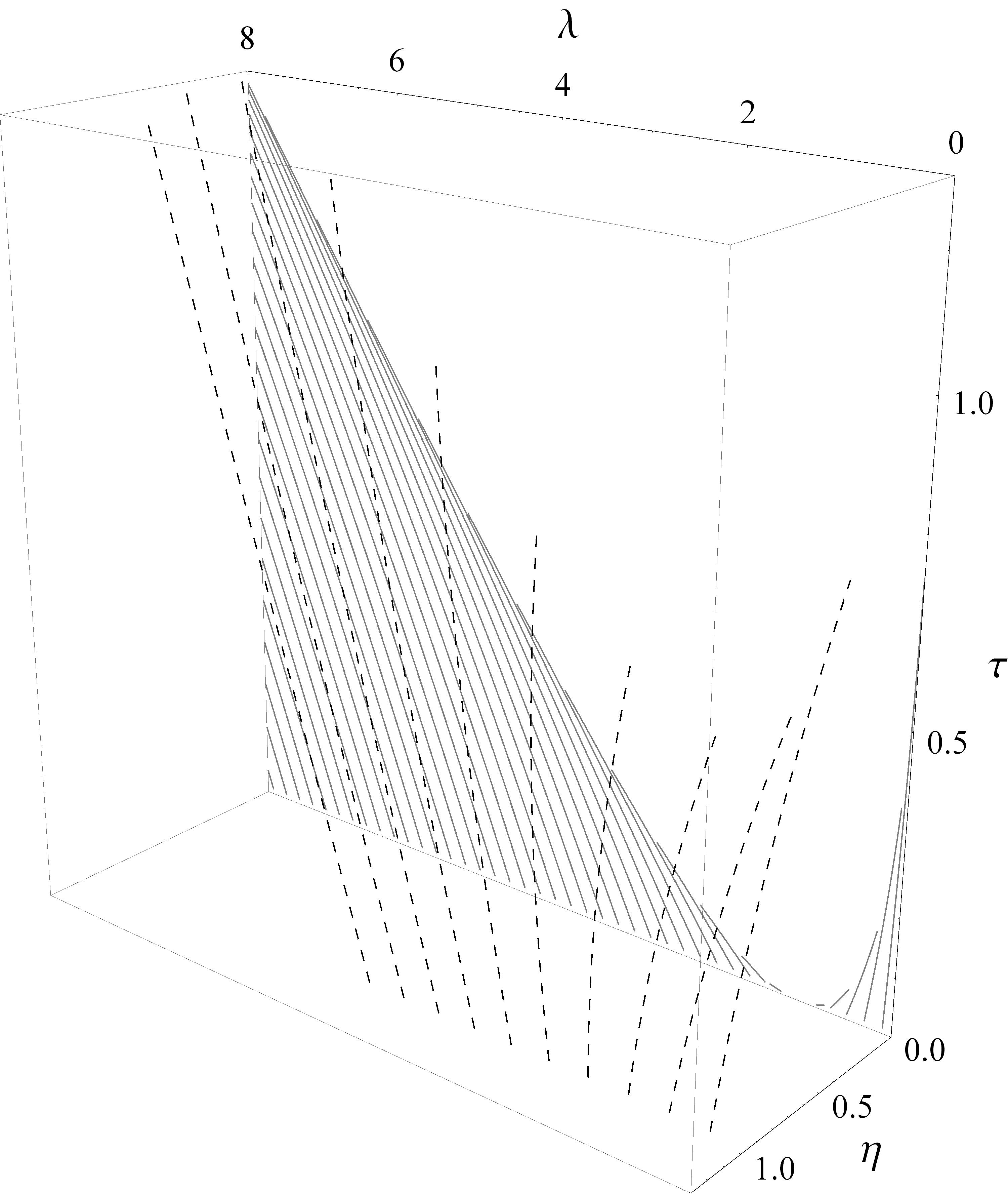}
		\caption{A sample of characteristic curves remaining real through the evolution is depicted with solid lines in the plane $\eta=0$. They cross each other on the edges of the large $N$ limit spectrum. 
The dashed lines are examples of characteristic curves which start at complex points ($\eta=1$). At a specific value of $\tau$, they cross the plane $\eta=0$
 in the area of the non-zero eigenvalue probability density.  }
		\label{charp}
	\end{figure}

The edge of the spectrum in the real $z$ plane is defined by $y=0$ and $Y(x,y)=0$ fulfilled simultaneously,  namely
\be
x^3-\tau(\tau+2)x-2\tau^2=0.
\ee
This  coincides with (\ref{shock1}). The main features of the characteristics described here are illustrated FIG.\ref{charp}.
	
\end{document}